\documentclass[]{aa}  

\usepackage{graphicx}
\usepackage{comment}    
\usepackage{xcolor}     
\bibpunct{(}{)}{;}{a}{}{,} 
\usepackage{natbib}
\usepackage[varg]{txfonts}
\begin{document}

   \title{Discovery of a 21 Myr old stellar population in the Orion complex}

   \author{Janez Kos
          \inst{1,2}
          \and
          Joss Bland-Hawthorn\inst{2,3}
          \and
          Martin Asplund\inst{3,4}
          \and
          Sven Buder\inst{5,6}
          \and
          Geraint F. Lewis\inst{2}
          \and
          Jane Lin\inst{4}
          \and
          Sarah L. Martell\inst{3,7}
          \and
          Melissa K. Ness\inst{8,9}
          \and
          Sanjib Sharma\inst{2}
          \and
          Gayandhi M. De Silva\inst{10}
          \and
          Jeffrey D. Simpson\inst{7}
          \and
          Daniel B. Zucker\inst{10,11}
          \and
          Toma\v{z} Zwitter\inst{1}
          \and
          Klemen \v{C}otar\inst{1}
          \and
          Lorenzo Spina\inst{12}
          }
    
   \institute{Faculty of mathematics and physics, University of Ljubljana, Jadranska 19, 1000 Ljubljana, Slovenia
   \\
   \email{janez.kos@fmf.uni-lj.si}
   \and
   Sydney Institute for Astronomy, The University of Sydney, School of Physics A28, Camperdown, NSW 2006, Australia
   \and
   Centre of Excellence for All Sky Astrophysics in 3D (ASTRO-3D), Australia
   \and
   Research School of Astronomy \& Astrophysics, Australian National University, ACT 2611, Australia
   \and
   Max Planck Institute for Astronomy (MPIA), Koenigstuhl 17, 69117 Heidelberg, Germany
    \and
    Fellow  of  the  International  Max  Planck  Research  School for  Astronomy  \&  Cosmic  Physics  at  the  University  of Heidelberg, Germany
    \and
    School of Physics, UNSW, Sydney, NSW 2052, Australia
    \and
    Department  of  Astronomy,  Columbia  University,  Pupin Physics Laboratories, New York, NY 10027, USA
    \and
    Center  for  Computational  Astrophysics,  Flatiron  Institute, 162 Fifth Avenue, New York, NY 10010, USA
    \and
    Department of Physics and Astronomy, Macquarie University, Sydney, NSW 2109, Australia
    \and
    Macquarie University Research Centre for Astronomy, Astrophysics and Astrophotonics, Macquarie University, Sydney, NSW 2109, Australia
    \and
    Monash Centre for Astrophysics, School of Physics and Astronomy, Monash University, VIC 3800 Australia}

   \date{Received September 31, 2018; accepted February 29, 2019}

 
  \abstract
   {The Orion complex is arguably the most studied star-forming region in the Galaxy. While stars are still being born in the Orion nebula, the oldest part was believed to be no more than 13 Myr old.}
   {In order to study the full hierarchy of star formation across the Orion complex, we perform a clustering analysis of the Ori~OB1a region using new stellar surveys and derive robust ages for each identified stellar aggregate.}
   {We use \textit{Gaia}~DR2 parameters supplemented with radial velocities from the GALAH and APOGEE surveys to perform clustering of the Ori~OB1a association. Five overdensities are resolved in a six-dimensional parameter space (positions, distance, proper motions, and radial velocity). Most correspond to previously known structures (ASCC~16, 25~Orionis, ASCC~20, ASCC~21). We use \textit{Gaia}~DR2, Pan-STARRS1 and 2MASS photometry to fit isochrones to the colour-magnitude diagrams of the identified clusters. The ages of the clusters can thus be measured with $\sim$10\% precision.}
   {While four of the clusters have ages between 11 and 13~Myr, the ASCC~20 cluster stands out at an age of 21$\pm 3$~Myr. This is significantly greater than the age of any previously known component of the Orion complex. To some degree, all clusters overlap in at least one of the six phase-space dimensions.}
   {We argue that the formation history of the Orion complex, and its relation to the Gould belt, must be reconsidered. A significant challenge in reconstructing the history of the Ori~OB1a association is to understand the impact of the newly discovered 21~Myr old population on the younger parts of the complex, including their formation.}


   \maketitle
%

\section{Introduction}

The Orion complex often serves as a proxy for a large star-forming region. Its proximity ($\sim400\, \mathrm{pc}$), size ($>10^4$ stars) and a mix of stars and gas at different evolutionary stages make the Orion complex an ideal place to study star formation and evolution, the destruction of young clusters and associations, the evolution of dense molecular clouds, and the interaction of the ISM with stars and supernova explosions \citep{reip08}. Indeed, much of what we understand today about star formation has come from detailed studies of the Orion complex \citep{krumholz19}.

The Orion complex forms part of the Gould belt \citep{poppel97}, a large ($\sim$1 kpc), young (30 - 40~Myr), ring-like structure tilted $\sim20^\circ$ to the Galactic plane that was discovered by Herschel in 1847. It can be observed as a concentration of young stars and molecular gas. Different scenarios for the formation of the Gould belt are found in the literature, like collisions of in-falling gas clumps from a Galactic fountain \citep{reip08} or a collision of one large high velocity cloud \citep{comeron94} with the local interstellar medium (ISM), possibly even a dark matter cloud \citep{bekki09}. A globular cluster could have acted in the same way \citep{bobylev18}. 

\begin{figure}
    \centering
    \includegraphics[width=\columnwidth]{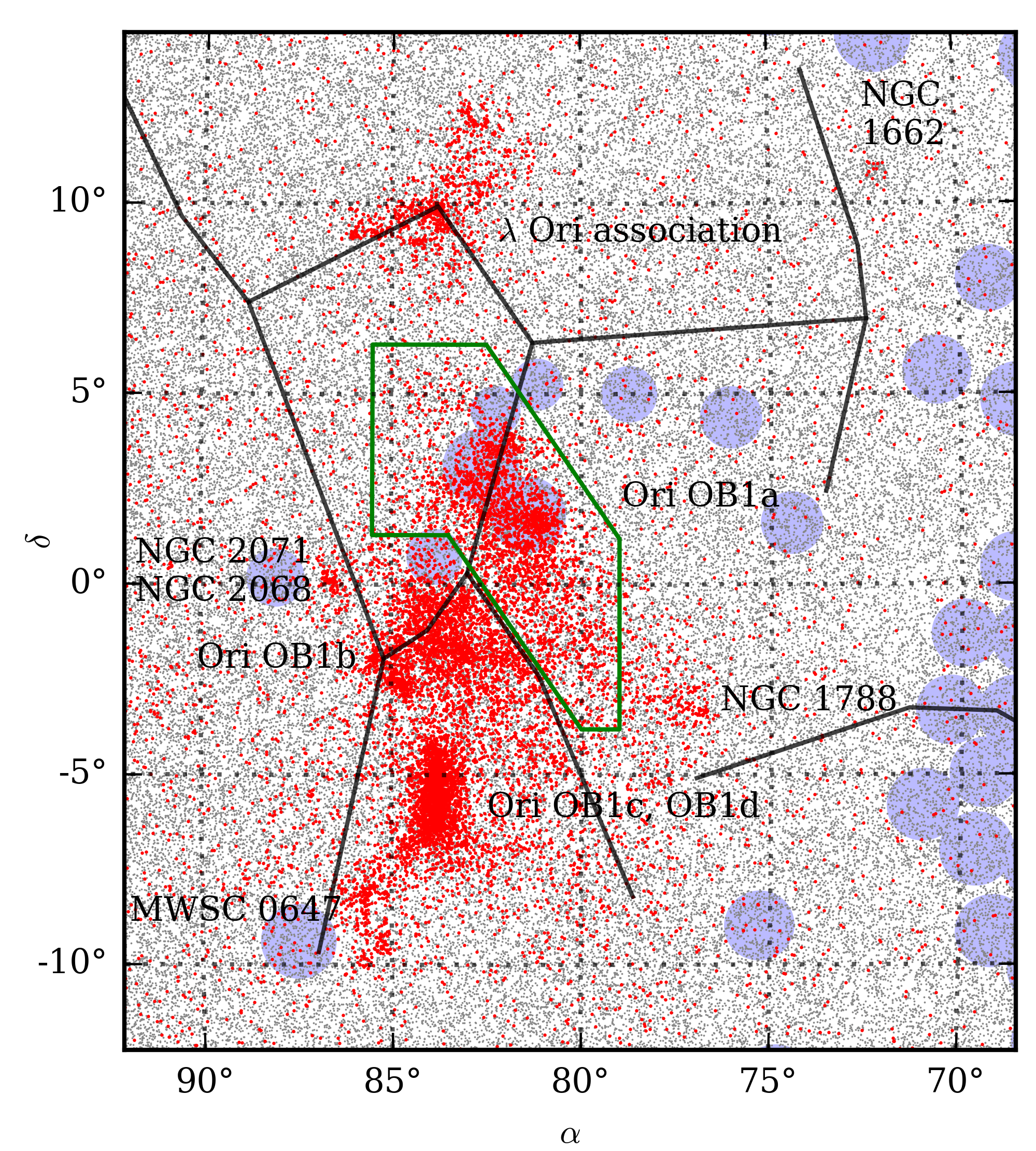}
    \caption{Star map of the the Orion complex. \textit{Gaia} DR2 stars in the Ori~OB1 distance range ($1.9<\varpi<3.5\ \mathrm{mas}$) are plotted in grey. Stars that also share Ori~OB1's proper motions ($-3.5<\mu<3.5\ \mathrm{mas\, yr^{-1}}$ for both components) are plotted in red. There are four main OB associations in Orion: $\lambda$~Ori association is centred on the star of the same name. Just above Orion's belt is Ori~OB1a, in Orion's belt is Ori~OB1b, around the Orion nebula is Ori~OB1c and the stars in the centre of the nebula form Ori~OB1d. Some open and embedded clusters associated with the complex are also marked. Fields observed in the GALAH survey are marked with blue circles. The Ori~OB1a region as used in this paper is marked with a green polygon. Black lines show the constellations Orion and Eridanus.}
    \label{fig:orion}
\end{figure}

The Orion complex shows a rich substructure (illustrated in Figure \ref{fig:orion}) and an age gradient, suggesting the star formation was triggered by an external event after the formation of the Gould belt and then spread throughout the region aided by feedback from the first-born stars \citep{lee05}. The Ori~OB1 association represents the main part of the complex and is further divided into four parts; above Orion's belt is Ori~OB1a, in Orion's belt is Ori~OB1b, around Orion nebula is Ori~OB1c and the stars in the centre of the nebula form Ori~OB1d \citep{blaauw64}. Ori~OB1a is the oldest region, until now believed to be up to 13~Myr old \citep{briceno07, downes2014, zari17, suarez17, kounkel18, briceno18} with an age spread of around 5~Myr (the value varies significantly in the literature, probably due to measurement uncertainties and pollution from younger parts of the association). The Ori~OB1a association shows several overdensities when projected on the sky \citep{kharchenko13}. The most prominent one is called the 25~Ori group, or ASCC~16 in \citet{kharchenko13}, which was until now believed to be the oldest structure in the complex \citep{zari17}, including a frequently discussed foreground population \citep{bouy14,fang17}. Ori~OB1a -- the main focus of this study lies at distances between 330 and 430~pc.

A meaningful separation of the Ori~OB1a association into distinct groups using clustering algorithms\footnote{By clustering we mean an algorithm able to resolve overdensities in a multi-dimensional space filled with points (stars). The overdensities are not necessary clusters of stars in a traditional sense, like open star clusters.}  has only become possible in the \textit{Gaia} era. The overdensities have been catalogued before \citep{kharchenko13, zari17}, but the clusters are superimposed so precise proper motions and distances are needed to disentangle them \citep{kounkel18}. Here we focus our effort on the oldest part of the complex -- Ori~OB1a only. To understand the early bursts of star formation in Ori~OB1a it is essential to resolve the oldest population of stars. In the past the age spread of Ori~OB1a has been measured, but not much effort has been put toward clustering the association and checking whether the age spread is due to continuous star formation or due to a few bursts. Precise \textit{Gaia} photometry also helps with measuring the ages of the clusters.

Triggered star formation is the prevailing theory of star formation in the Orion complex \citep{cunha}. With the advance of massive spectroscopic surveys measuring chemical abundances of a large number of stars and covering all structures in the complex, it could also be established whether the older populations of stars chemically enriched the younger ones. This is supported by the chemical gradients observed in the complex \citep{orazi09, biazzo11}, but a complete hierarchy inside the complex will have to be resolved, including the recent history and dynamical interactions, before this can be proved. Our work shows that basic building blocks can be resolved in the Orion complex and a careful clustering analysis can reveal structures that are extremely important for the understanding of the history of the Orion complex and were missed in past research.

\section{Data}

The \textit{Gaia}~DR2 \citep{gaia18} catalogue is complete between $G$=12 and 17 and almost complete at brighter magnitudes. Only a few of the brightest Ori~OB1 stars are missing. While they would improve the analysis presented here, there is no practical way to include them, as no photometric measurement exists for them in \textit{Gaia} bands. \textit{Gaia}~DR2 includes qualitatively and quantitatively unprecedented photometry. $G$ magnitudes have an uncertainty of $<0.001$ (statistical), $<0.01$ (systematic) magnitude for almost all stars with $G$<17 and the uncertainty of the BP-RP colour is $<0.01$ for almost all stars with $G$<16 \citep{evans18}. In this work we also used Pan-STARRS1 \citep{flewelling16} magnitudes. Their uncertainty is around 0.01 mag for virtually all stars used in this work \citep{magnier13}. 2MASS was used as well and the uncertainty for the stars used in this work is $<0.03$ mag \citep{skrutskie06}. Stars in the \textit{Gaia}~DR2 catalogue have been cross-matched with Pan-STARRS1 and 2MASS using \textsc{Xmatch} by CDS. More than 90\% of stars are found in all three catalogues. Only two quality cuts were used in the selection of \textit{Gaia} data: $\sigma_\mu<0.35\ \mathrm{mas\, yr^{-1}}$ and $\sigma_\varpi<0.2\ \mathrm{mas}$.

\textit{Gaia} radial velocities are only available for 7\% of Ori~OB1a stars with an average uncertainty of $4.9\  \mathrm{km\, s^{-1}}$. This is enough to resolve members of the association from the field stars. It is, however, more than an order of magnitude worse than the precision of \textit{Gaia}'s proper motions. For the benefit of clustering, we supplemented the \textit{Gaia} radial velocities with GALAH~DR2 radial velocities \citep{buder18, zwitter18} and APOGEE DR14 radial velocities \citep{abolfathi18}, where available. Both spectroscopic surveys have a narrow magnitude selection function, so radial velocities are expected to exist only for a limited number of stars. However, because the surveys also observe stars with no \textit{Gaia} radial velocities, they both supplement and/or improve \textit{Gaia}'s radial velocities. 56 stars used in this work have only GALAH radial velocities and additional 53 have APOGEE radial velocities. They are much more precise than \textit{Gaia's} at a typical uncertainty of $0.3\ \mathrm{km\, s^{-1}}$.

There is no strict definition of Ori~OB1a, as the components of the Orion complex are superimposed onto each other. We define the region of interest by a polygon described by vertices:
\begin{multline}
(\alpha, \delta)_{\mathrm{J}2000.0}=\big \{ (83.5^\circ, 1.3^\circ),\ (85.5^\circ, 1.3^\circ),\ (85.5^\circ, 6.3^\circ),\\
(82.5^\circ, 6.3^\circ),\ (79.0^\circ, 1.2^\circ),\ (79.0^\circ, -3.8^\circ),\ (80.0^\circ, -3.8^\circ) \big \}.
\end{multline}

\section{Clustering O\lowercase{ri} OB1\lowercase{a}}

\begin{figure*}
    \definecolor{c_r}{HTML}{b00000}
    \definecolor{c_r1}{HTML}{ff5500}
    \definecolor{c_r2}{HTML}{ff006a}
    \definecolor{c_r3}{HTML}{ffd100}
    \definecolor{c_b}{HTML}{0000b0}
    \definecolor{c_b1}{HTML}{6600ff}
    \definecolor{c_b2}{HTML}{009999}
    \centering
    \includegraphics[width=0.99\textwidth]{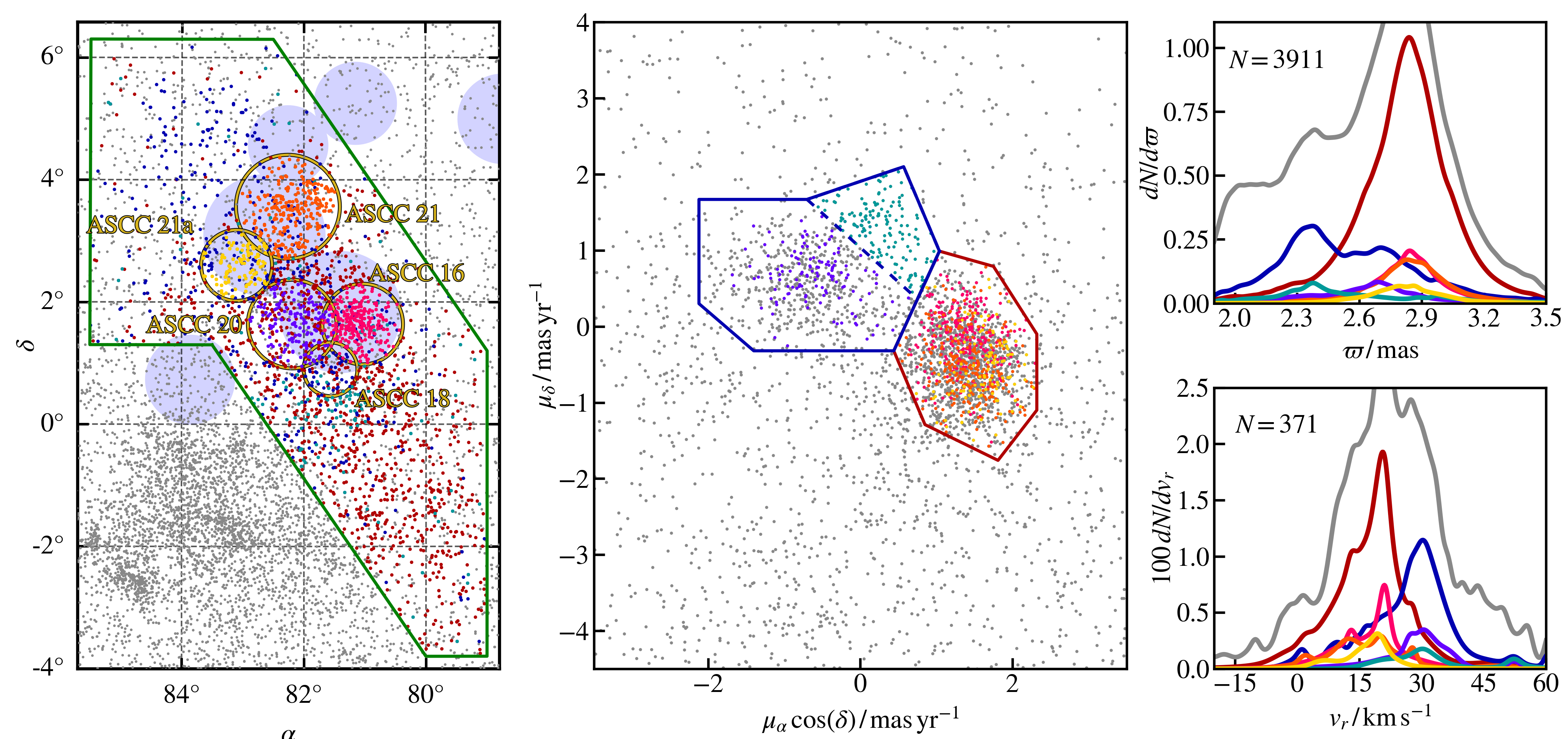}
    \includegraphics[width=0.99\textwidth]{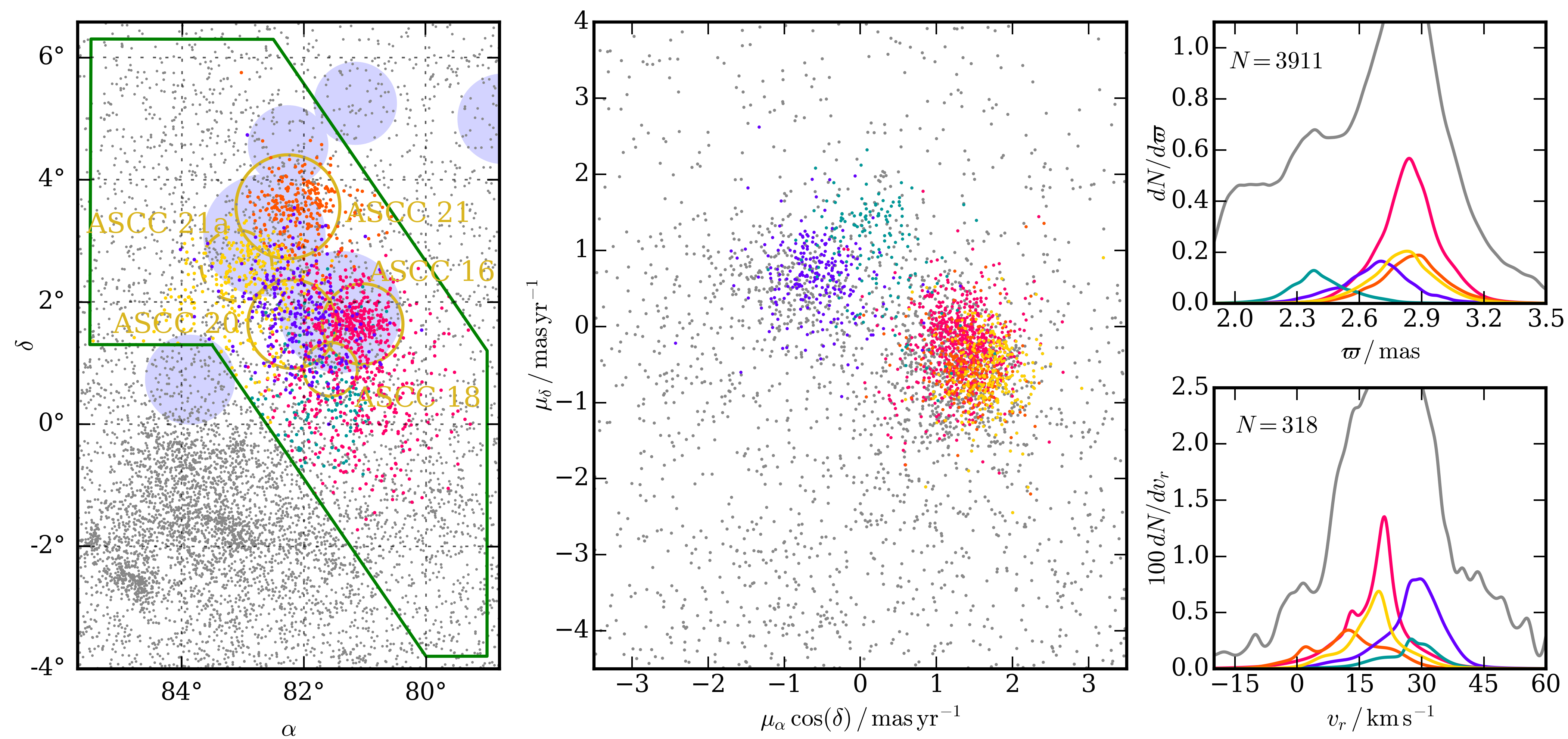}
    \includegraphics[width=0.99\textwidth]{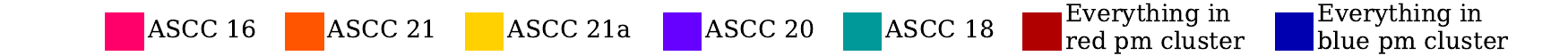}
    \caption{Clustering of Ori~OB1a. Top panels show the initial heuristic clusters and bottom panels show the final clustering. Left: Star map of the association. Only stars between $-3.5<\mu_\alpha \cos{(\delta)}<3.5\, \mathrm{mas\, yr^{-1}}$, $-4.5<\mu_\delta<4\, \mathrm{mas\, yr^{-1}}$, and $1.9<\varpi<3.5\, \mathrm{mas}$ are plotted. Yellow circles mark associations in \citet{kharchenko13}. The circle marked as ASCC 21a was added by us. Blue filled circles mark fields observed in the GALAH survey. Middle: The proper motion plane. Only stars inside the green polygon from the left-hand panel are plotted. Red and blue polygons mark two distinct proper motion groups that we use in heuristic clustering. Top-right: Distribution of stars' parallaxes inside the green polygon (grey line). Bottom-right: Distribution of stars' radial velocities inside the green polygon (grey line). Only a fraction of stars have radial velocity measured in \textit{Gaia}~DR2. For stars observed in the GALAH and APOGEE surveys, we use the radial velocities measured by them. Distributions for individual groups are shown with the matching colours. 
    }
    \label{fig:ob1a}
\end{figure*}

Even at first glance obvious substructures are visible in the Ori~OB1a association. Our goal was to cluster the Ori~OB1a stars into groups that represent real building blocks of the association and are not just apparent overdensities with no physical relation between stars. Different parts of the association significantly overlap in the 6D space ($\alpha$, $\delta$, $\mu_\alpha$, $\mu_\delta$, $\varpi$, $v_r$). This is partially due to measurement uncertainties, but mostly due to the dynamical processes and the nature of the star-forming region. As a consequence, an automatic clustering analysis is difficult. We tried several clustering methods with very limited success. Agglomerative clustering \citep{pedregosa11} was not able to take the background/field stars into the account. All clusters included a large fraction of stars which were classified as non-members in the HR diagram (photometry was not used in the clustering process but is used for verification). Hierarchical DBSCAN \citep{mcinnes17, mcinnes17b} had the opposite problem. With only 5 clusters, most stars were classified as background. Many more smaller clusters were needed to include most Ori OB1a stars. Robust single linkage \citep{ mcinnes17b} from the same Python module as the Hierarchical DBSCAN was less susceptible to the last problem, but still not suitable. Sizes of clusters produced with the three mentioned methods also varied too much in size, were unstable to small parameter variations, and sometimes had unnatural shapes. The only one that produced reasonable results was ENLINK \citep{sharma09}. However, we use a simpler approach here, as ENLINK could not readily be modified to accommodate a custom metric and include radial velocities. Our simple procedure is able to replicate ENLINK clustering, but requires manual initialisation. 

Clusters were first found heuristically by selecting different regions in the 6D space until a number of clearly separated, oval-shaped clusters were found. How the 6D space was partitioned and the clusters we found in this step are shown in Figure~\ref{fig:ob1a}, top panels. In the $(\alpha,\delta)$ plane we used regions marked as clusters ASCC~16, ASCC~18, ASCC~20, and ASCC~21 in \citet{kharchenko13} and added a region marked as ASCC~21a where we observed another overdensity.

Such heuristic clusters are very unnatural -- they were not allowed to overlap (at least not in all dimensions) and have subjectively determined boundaries. We fixed this in the next step by finding the nearest cluster centre for each observed star. A custom metric for measuring distances between stars and cluster centres in 5D space is:
\begin{equation}
\label{eq:met}
    d=\frac{\arccos{\left( \mathbf{r} \cdot \mathbf{\overline{r}} \right)}}{1.25^\circ}+\frac{\sqrt{\left( \mu_\alpha - \overline{\mu_\alpha} \right)^2+\left(\mu_\delta - \overline{\mu_\delta} \right)^2}}{1.0\, \mathrm{mas\, yr^{-1}}}+\frac{\left | \varpi-\overline{\varpi} \right |}{0.22\, \mathrm{mas}},
\end{equation}
where $\mathbf{r}=\left(\alpha, \delta \right)$.\\
The first term is a great circle distance between two points on the sky, the second term is distance in the proper motion plane, and the third term is an absolute difference in parallaxes between two stars. Bars denote a mean value for a cluster (cluster centre). All three terms are made non-dimensional and have weights applied to them, so each dimensionless term is in the same order of magnitude for Ori~OB1a stars. The values of the weights are not critical and do not have to be fine-tuned. All stars within $d<3.1$ of at least one cluster are re-labelled as members of the nearest cluster. Unlike the heuristic clustering, clusters can now overlap in any dimension and only the maximum size of the cluster ($d=3.1$) is artificially introduced. $d=3.1$ was chosen, so the clusters are as large as possible without introducing too much pollution from field stars.

One can see from the radial velocity distribution in Figure~\ref{fig:ob1a}, second panel from the top, that the clustering would benefit if the radial velocity were taken into account, where possible. The metric must be modified to include that:
\begin{equation}
\label{eq:met2}
    d_{\mathrm{6D}}=\frac{3}{4}\left (d+\frac{\left | v_r-\overline{v_r} \right |}{15.0\, \mathrm{km\,s^{-1}}}\right ).
\end{equation}
The $3/4$ factor makes the distances with or without the radial velocity term comparable, because a distance is always larger, if more terms are added. Since all four terms are normalised to the same dimensional scale, the distance with four terms would be 25\% larger in average.

\begin{table*}
    \centering
    \begin{tabular}{ccccccccccc}
    \hline\hline
    Cluster & $\alpha$ & $\delta$ & $r$ & $\mu_\alpha\, \cos{\delta}$ & $\mu_\delta$ & $\varpi$ & $v_r$ & $\sigma_v$ & Age & N. of stars\\[3pt]
     & $^\circ$ & $^\circ$ & $^\circ$ & $\mathrm{mas\, yr^{-1}}$ & $\mathrm{mas\, yr^{-1}}$ & $\mathrm{mas}$ & $\mathrm{km\, s^{-1}}$ & $\mathrm{km\, s^{-1}}$ & Myr & (with $v_r$)\\\hline
     ASCC 16 & 81.171 & 1.386 & 0.87 & 1.33 & -0.23 & 2.84 & 19.72 & 6.45 & 13.0$\pm1.3$ & 683 (50)\\
ASCC 18 & 81.569 & 0.427 & 0.74 & 0.20 & 1.16 & 2.41 & 27.19 & 5.77 & 12.75$\pm1.27$ & 148 (11)\\
ASCC 20 & 82.135 & 1.792 & 0.80 & -0.56 & 0.67 & 2.69 & 27.70 & 7.66 & 21.25$\pm2.12$ & 237 (41)\\
ASCC 21 & 82.040 & 3.530 & 0.54 & 1.40 & -0.55 & 2.87 & 19.12 & 5.98 & 11.0$\pm1.1$ & 266 (46)\\
ASCC 21a & 82.904 & 2.297 & 0.72 & 1.70 & -0.65 & 2.80 & 19.50 & 7.38 & 12.75$\pm1.27$ & 282 (39)\\
    \hline
    
    \end{tabular}
    \caption{Parameters used in the final step of the clustering algorithm ($\alpha$, $\delta$, $\mu_\alpha\, \cos{\delta}$, $\mu_\delta$, $\varpi$, $v_r$) and some derived quantities ($r$ -- the radius enclosing 50\% of the stars, $\sigma_v$ -- the dispersion of the radial velocity, and age). Number of stars gives the total number and the number with known radial velocities in the brackets.}
    \label{tab:parameters}
\end{table*}

After the last step the cluster centres were recalculated and a few iterations of the above process were made. This step is similar to the K-Means algorithm, but allows for unclustered stars and forces the cluster to be fairly circular in 6D space. Final clusters are shown in Figure~\ref{fig:ob1a}, bottom panels. Parameters of the final five clusters are collected in Table \ref{tab:parameters}. Note that the cluster ASCC~18 is shifted with respect to the values in \citet{kharchenko13}.

We plotted the stars from all five clusters on an HR diagram (Figure~\ref{fig:HR_fitted}) to verify the accuracy of the clustering method. Ori~OB1a stars are expected to lie on a pre-main sequence (PMS), so any star on the main sequence is a mis-identified cluster member. We estimate that less than 5\% of stars were mis-identified. All stars more than 0.1 magnitude away from the fitted isochrone (or the binary sequence) on a \textit{Gaia} HR diagram were rejected as mis-identified field stars and are not shown in Figure~\ref{fig:HR_fitted}.

\section{Cluster ages}

\begin{figure*}
    \centering
    \includegraphics[width=\textwidth]{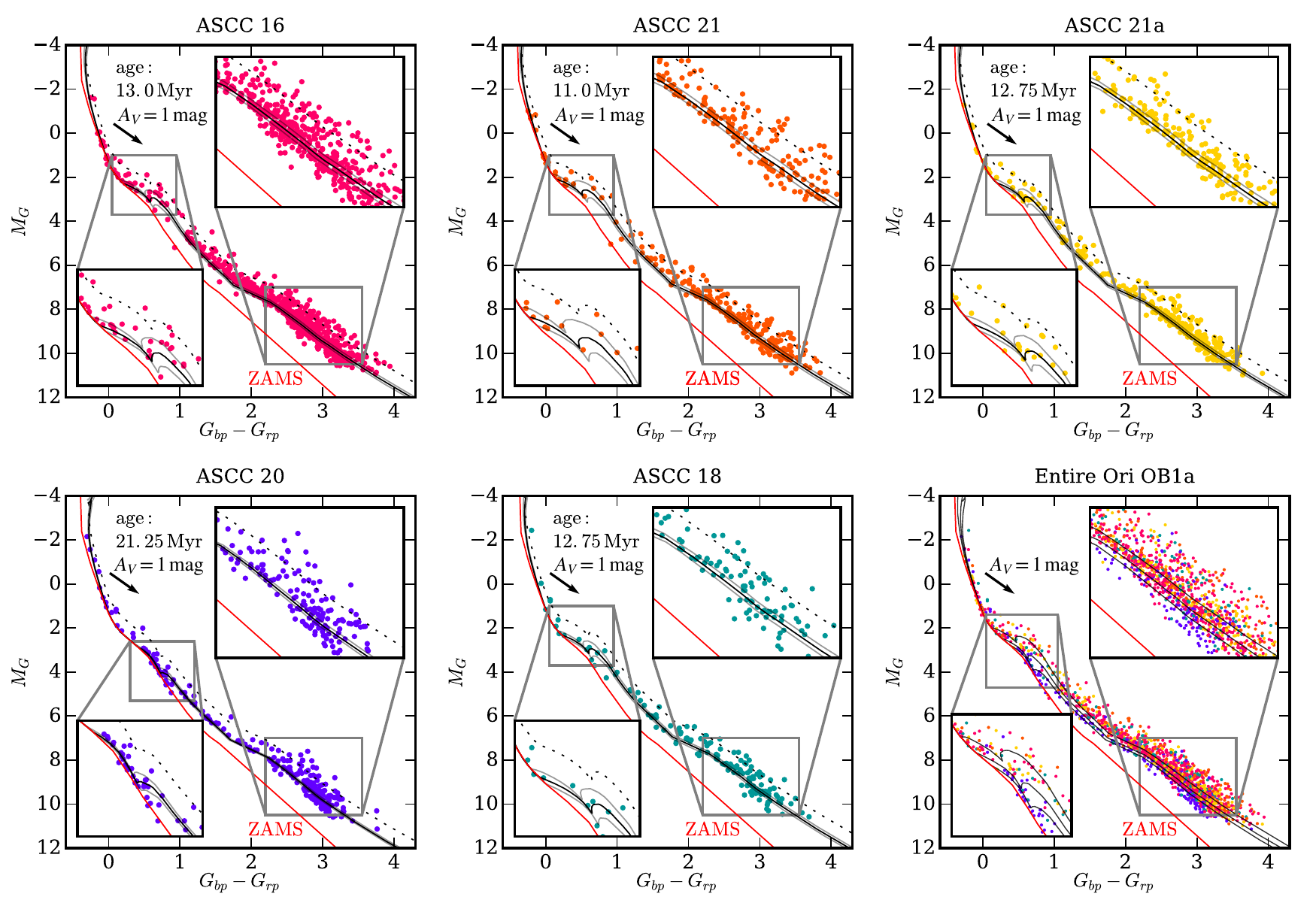}
    \caption{Isochrones fitted to the \textit{Gaia}~DR2 HR diagrams. Each panel shows one group in Ori~OB1a. Solid lines are the best matching isochrones. Dashed lines are the isochrones for the binary sequence (for stars of equal luminosities). The age of the best matching isochrone is written in each panel. Isochrones for $\pm2\, \mathrm{Myr}$ are plotted in gray. The zero-age-main-sequence is plotted in red. For all five groups we assumed the same metallicity $\left(\left[ \mathrm{M/H} \right]=0.03\right)$ and extinction ($A_V=0.25$). Zoomed insets show the regions most suitable for age determination. The last panel shows all five clusters together with isochrones for 10, 15, and 20~Myr. The arrow on each plot shows a reddening vector at $A_V=1$~mag. See also an online animation showing the comparison of the HR diagrams. Note that the darkest stars plotted here at $G\sim17.5$ are bright enough to be unaffected by background subtraction issues in Gaia DR2.}
    \label{fig:HR_fitted}
\end{figure*}

We fitted PARSEC isochrones \citep{bressan12, chen14, tang14} to \textit{Gaia}~DR2, Pan-STARRS1 \citep{flewelling16}, and 2MASS \citep{skrutskie06} magnitudes to obtain the age, photometric metallicity and extinction of each cluster. PARSEC isochrones were used because they are reliable over the whole HR diagram covered in this work. See \citet{herczeg15} for comparison with other isochrones. Age, metallicity and extinction are too degenerate to be fitted individually to each cluster, so we assumed the same metallicity ([M/H]=0.03) and extinction ($A_V=0.25$) for all five clusters in order to constrain the individual ages. The values of metallicity and extinction were obtained by fitting isochrones with similar metallicity and extinction as found in the literature \citep[e.g.][]{orazi09, kharchenko13, green18}. The best matching isochrones are plotted in Figure~\ref{fig:HR_fitted} using \textit{Gaia} magnitudes. Most Ori~OB1 stars are PMS stars, so the age can be best constrained from a strong correlation between age and distance to the zero age main sequence (ZAMS) for K and M dwarfs, and the point where the PMS merges with the ZAMS. Both regions are enhanced in Figure~\ref{fig:HR_fitted}. There are fewer stars in the latter region, but the isochrones are not as well established in the former region. The region where PMS isochrones touch the ZAMS is therefore most useful for constraining the absolute age, and the K and M dwarfs can be used to identify differential age between five clusters. The isochrones have been fitted by eye and we claim a statistical age uncertainty of $\sim10$\%. Appendix \ref{sec:uncertainty} discusses age uncertainty estimation, which can be either estimated when fitting by eye or statistically estimated in the lower main sequence region. Stars on the ZAMS (above $M_G=2$) can be used to constrain extinction and metallicity.

We can clearly distinguish a significantly older age of ASCC~20, but also detect an age difference of 2~Myr between ASCC~21 and the other three clusters. The age determination could be improved, if unresolved binary stars (because they have a brightness offset proportional to the luminosity excess compared to a single star) and fast rotating stars (because their luminosity depends on rotational velocity and inclination \citep{maeder00, georgy13}) were identified in the region where the PMS merges with the ZAMS. This can be done from spectroscopic observations, but spectra are available only for a small number of stars. 

Within individual clusters we observe no age spread that could not be explained by measurement uncertainties (in addition to the unresolved binaries and fast rotators discussed above). For \textit{Gaia} and Pan-STARRS1 magnitudes it is due to parallax uncertainties (they translate into $\sim0.08$ absolute magnitude error) and for 2MASS magnitudes the actual magnitude uncertainty is also important.

\section{Discussion}

Differential extinction in Ori~OB1a is relatively low ($\sigma_{A_V}=0.05$) \citep{green18}, allowing precise age determination without correcting the magnitudes of individual stars for their exact extinction (which is rarely well known). The extinction also increases with distance, but the difference between the nearest clusters (ASCC~16, ASCC~21, ASCC~21a) and the farthest cluster (ASCC~18) would be $\Delta A_V=0.06$, if the extinction uniformly increased inside the association as 3D extinction maps show \citep{green14, green18, lombardi11, capitano17}. We did not take this into the account when calculating ages, but a difference in age would be $<0.25\, \mathrm{Myr}$. Systematic uncertainties in parallax zero point (around 0.04 mas \citep[e.g.][]{lindegren18}) mean or 0.035 uncertainty in $M_G$ magnitude, which again translates into $<0.25\, \mathrm{Myr}$.

There are two reasons we use an unorthodox metric shown in Equations \ref{eq:met} and \ref{eq:met2}. Quantities we use have different uncertainties and consequently different scatter in each dimension (Almost no uncertainty in position on the sky vs. large uncertainty in distance, for example). Manhattan distances are less sensitive to this than Euclidean distances. This is more pronounced the more dimensions there are. The parameter space becomes emptier and all distances become large when more dimensions are used. Manhattan (or $L_1$) metric is most sensitive to differences in distances between scattered points of any $L_p$ metrics. We are also not trying to quantify the multivariate scatter, so there is no need to add the terms of Equations \ref{eq:met} and \ref{eq:met2} in quadrature. The goal instead  is to best resolve clusters.

Clustering done by \citet{kounkel18} and \citet{briceno18} does include ASCC~20 (called HR~1833 in their works), but their ages do not push past the previously established age of around 13~Myr for oldest structures in the Orion complex. Ages reported in this work do not match well with \citet{kounkel18} ages (two methods give $15.1\pm3.4$ and $12.9\pm2.8$ Myr for ASCC~20, and between 4.9 and 7.4 Myr for other clusters) due to vastly different age determination methods. However, our ages do agree with \citet{bossini19} ages for ASCC~21 ($10.9\pm0.3$ Myr) and ASCC~16 ($11.2\pm0.1$ Myr) as measured by Bayesian fitting of isochrones. Other three clusters were not measured in \citet{bossini19}.

ASCC 20 pushes the age of the Orion complex closer to the estimated age of the Gould belt, but does not exceed the lowest age of the Gould belt found in the literature \citep{perrot03}. ASCC~20 does not stand out only due to its age. Its kinematics are peculiar in respect to most stars in Ori~OB1a. It is slightly more distant than most ORi~OB1a stars, but its velocity shows that it used to be closer to the rest of the Ori~OB1a stars a few million years ago. It is likely the ASCC~20 stars were born away from other clusters (although still inside the Orion complex) and interacted with them later. Orbit integration of stars inside the associations is beyond the scope of this paper and we plan to study the mentioned interaction in future work together with immense spectroscopic information from surveys like APOGEE \citep{cottle18} and GALAH \citep{buder18}.

Our clustering algorithm has a built-in maximum size of the cluster in the six dimensional space. Since the maximum size must be low enough to prevent pollution from field stars or from other associations nearby, it can also exclude possible members of the clusters in question that could be detected with other methods, either by a different clustering method or by using a different parameter space. Some members, for example, could be found far from the cluster centre due to a peculiar velocity \citep{kos18}. Using orbital parameters in place of the six dimensional parameter space could solve this, but due to a lack of radial velocity measurements it can be done only for a small sub-sample of stars, statistically too small to provide reliable ages further in the analysis.

Our highly supervised clustering worked extremely well in a crowded region of Ori~OB1a. Any degeneracy between the clusters is well resolved, so only a few stars might have wrongly assigned clusters. On top of that the pollution from field stars is extremely low, which we verified with the initial HR diagrams.

\section*{Acknowledgements}

JK, TZ, and K\v{C} acknowledge the financial support from the Slovenian Research Agency (research core funding No. P1-0188). SLM acknowledges support from the Australian Research Council through Discovery Project 180101791. This research was supported by the Australian Research Council Centre of Excellence for All Sky Astrophysics in 3 Dimensions (ASTRO 3D), through project number CE170100013.




\bibliographystyle{aa}
\bibliography{bib}

\begin{appendix}
\section{Age uncertainty determination}
\label{sec:uncertainty}

Uncertainty estimate is hard to calculate when fitting by eye. While we did estimate the age uncertainty to be around 10\% by eye, we show a more quantitative approach here. The best fitting isochrone was determined by eye and here we assume it does indeed best fit the data, even-though Figure \ref{fig:error_calc} might not show this. We can define a distance from the isochrone to each star. As this is poorly defined on a colour-magnitude diagram, we define it on a magnitude-magnitude diagram (or in a magnitude space only). This can be done using any number of filters $f_1$ to $f_n$:

\begin{figure*}[!ht]
    \centering
    \includegraphics[width=\textwidth]{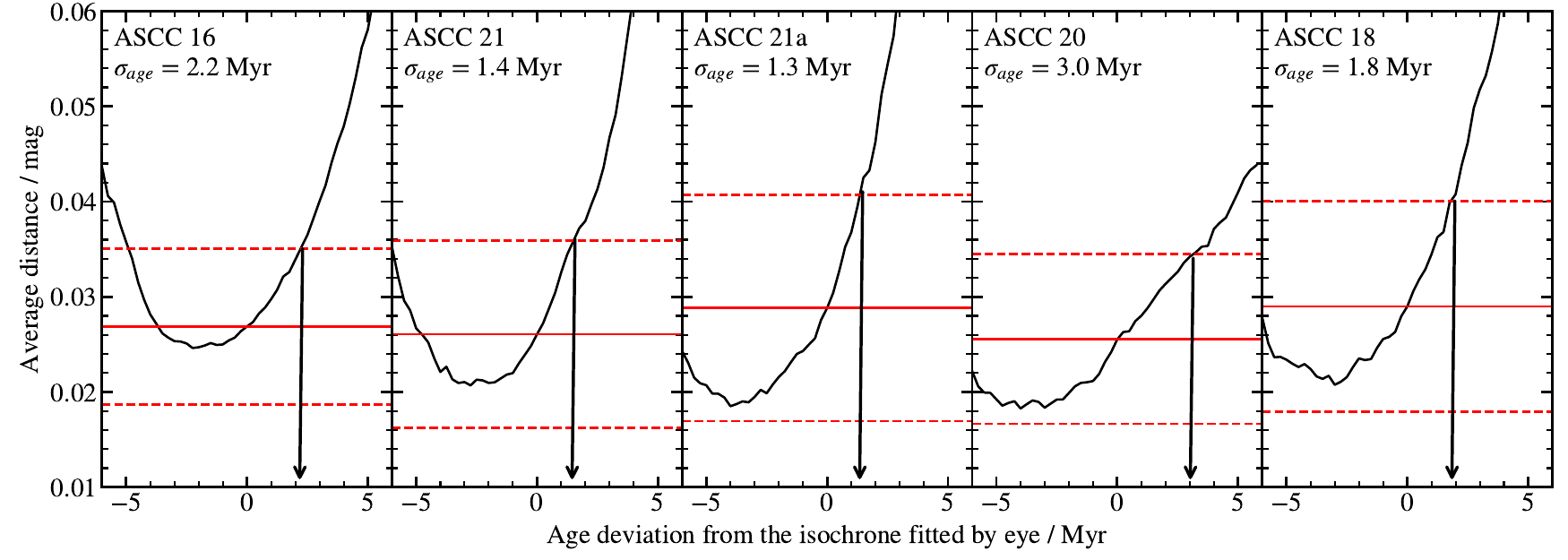}
    \caption{Derivation of age uncertainties. Black line shows average distance between all cluster members and an isochrone at a given age as measured on an HR diagram. Red horizontal lines show the average distance at our claimed age (solid line) and $1\sigma$ scatter of distances from the isochrone fitted by eye (dashed line). Age difference at which the average distance reaches the top dashed line is $1\sigma$ uncertainty for the age of each cluster. The calculated uncertainty is marked by an arrow. Note that such analysis is highly impacted by binaries and biases, so the described procedure is only reliable for positive age deviations.}
    \label{fig:error_calc}
\end{figure*}

\begin{equation}
d=d_\perp \left( I(f_1,f_2,\ldots, f_n), S_{f_1, f_2,\ldots,f_n} \right),
\end{equation}

where $d_\perp$ is the distance between a star $S$ and the nearest point on the isochrone $I$. Isochrone is a continuous function in a $n$ dimensional space. 

From distances to the isochrone for all cluster members we define an average distance of the cluster to the isochrone and some scatter around the isochrone (standard deviation of distances for all members). The scatter measured for the isochrone fitted by eye tells us how coherent the HR diagram is for this cluster. The average distance tells us how well does the isochrone fit the data -- or at least it would in an ideal case. Figure \ref{fig:error_calc} shows the average distance plotted for isochrones at different ages. Zero marks the isochrone we fitted by eye. One can see that younger isochrones have smaller average distance. This is due to binaries and systematic biases (like distance uncertainties) that position more stars above the isochrone, meaning toward younger ages. This is therefore not a reliable method to fit ages when biases are not well understood and properly taken into the account. In our case the ages would be underestimated by $\sim3.5$~Myr, if the best fitting isochrone was used for age. However, this approach can be used to estimate the sensitivity of isochrone fitting to age. 

On figure \ref{fig:error_calc} we also plotted the average distance of cluster members from the isochrone fitted by eye (solid red line) and the scatter of distances (dashed lines). Age at which the average distance becomes larger than the scatter is $1\sigma$ uncertainty for the cluster age. Only the upper uncertainty can be measured this way, so we can only assume the uncertainty is symmetric around the reported age. Average distance for younger isochrones is obviously highly impacted by the binaries and biases and no valid number can be deduced from the shown plots for the lower uncertainty.

A typical age uncertainty we measure is 13\%. The result is almost exclusively determined by cooler dwarfs on the PMS. When fitting by eye we also payed attention to the merging point between the PMS and the main sequence, so a 10\% uncertainty estimated by eye probably holds.

\section{List of members}
\longtab[1]{
\tiny
\def\arraystretch{0.75}
\begin{longtable}{lcccccc}
\caption{Members of ASCC 16.}\\
\hline\hline
\textit{Gaia} DR2 id & $\alpha$ & $\delta$ & $\mu_\alpha\, \cos(\delta)$ & $\mu_\delta$ & $\varpi$ & $v_r$ \\
& $^\circ$ & $^\circ$ & $\mathrm{mas\, yr^{-1}}$ & $\mathrm{mas\, yr^{-1}}$ & $\mathrm{mas}$ & $\mathrm{km\, s^{-1}}$ \\
\hline
\endfirsthead
\caption{Continued.}\\
\hline\hline
\textit{Gaia} DR2 id & $\alpha$ & $\delta$ & $\mu_\alpha\, \cos(\delta)$ & $\mu_\delta$ & $\varpi$ & $v_r$ \\
& $^\circ$ & $^\circ$ & $\mathrm{mas\, yr^{-1}}$ & $\mathrm{mas\, yr^{-1}}$ & $\mathrm{mas}$ & $\mathrm{km\, s^{-1}}$ \\
\hline
\endhead
\hline
\endfoot
3220166062483977088 & 81.12571 & -1.72150 & 1.233$\pm$0.046 & -0.288$\pm$0.039 & 2.845$\pm$0.028 &  \\
3220450629837013120 & 80.37219 & -1.11714 & 1.123$\pm$0.078 & -0.026$\pm$0.064 & 2.915$\pm$0.046 &  \\
3220462655745525632 & 80.87922 & -1.07325 & 1.269$\pm$0.091 & -0.516$\pm$0.066 & 2.821$\pm$0.048 &  \\
3220422381836144768 & 81.08407 & -1.04995 & 1.876$\pm$0.153 & -0.479$\pm$0.120 & 2.783$\pm$0.099 &  \\
3220306799971283200 & 81.81931 & -1.12785 & 1.711$\pm$0.047 & -0.257$\pm$0.043 & 2.802$\pm$0.030 &  \\
3220439119325046784 & 80.55365 & -1.23116 & 1.042$\pm$0.195 & -0.282$\pm$0.165 & 2.843$\pm$0.102 &  \\
3220453752276755200 & 80.53331 & -1.09230 & 1.480$\pm$0.047 & -0.105$\pm$0.037 & 2.871$\pm$0.025 &  \\
3220382082159267072 & 80.84557 & -1.44094 & 1.171$\pm$0.090 & -0.512$\pm$0.076 & 2.852$\pm$0.047 &  \\
\hline
\end{longtable}
}
\longtab[2]{
\tiny
\def\arraystretch{0.75}
\begin{longtable}{lcccccc}
\caption{Members of ASCC 21.}\\
\hline\hline
\textit{Gaia} DR2 id & $\alpha$ & $\delta$ & $\mu_\alpha\, \cos(\delta)$ & $\mu_\delta$ & $\varpi$ & $v_r$ \\
& $^\circ$ & $^\circ$ & $\mathrm{mas\, yr^{-1}}$ & $\mathrm{mas\, yr^{-1}}$ & $\mathrm{mas}$ & $\mathrm{km\, s^{-1}}$ \\
\hline
\endfirsthead
\caption{Continued.}\\
\hline\hline
\textit{Gaia} DR2 id & $\alpha$ & $\delta$ & $\mu_\alpha\, \cos(\delta)$ & $\mu_\delta$ & $\varpi$ & $v_r$ \\
& $^\circ$ & $^\circ$ & $\mathrm{mas\, yr^{-1}}$ & $\mathrm{mas\, yr^{-1}}$ & $\mathrm{mas}$ & $\mathrm{km\, s^{-1}}$ \\
\hline
\endhead
\hline
\endfoot
3234356733213504896 & 81.17837 & 2.58250 & 1.432$\pm$0.133 & -0.796$\pm$0.117 & 3.024$\pm$0.082 &  \\
3234634325541651840 & 80.63855 & 3.13454 & 1.065$\pm$0.217 & -0.791$\pm$0.169 & 2.787$\pm$0.130 &  \\
3235930825842491392 & 80.93914 & 3.04084 & 1.295$\pm$0.187 & -1.061$\pm$0.163 & 2.895$\pm$0.086 &  \\
3235906494852614656 & 81.87817 & 3.24829 & 0.669$\pm$0.301 & -0.939$\pm$0.264 & 3.045$\pm$0.244 &  \\
3235861517954827776 & 81.44415 & 2.78185 & 0.614$\pm$0.282 & -0.287$\pm$0.235 & 2.861$\pm$0.182 &  \\
3235922785663970304 & 81.45747 & 3.04622 & 1.280$\pm$0.214 & -0.126$\pm$0.175 & 3.136$\pm$0.134 &  \\
3235915913716183680 & 81.37971 & 2.88339 & 1.071$\pm$0.111 & -0.321$\pm$0.100 & 2.773$\pm$0.079 & 14.91$\pm$1.87 \\
3235861968926969472 & 81.31150 & 2.75368 & 1.322$\pm$0.150 & -0.410$\pm$0.122 & 2.820$\pm$0.084 & 4.61$\pm$7.46 \\
\hline
\end{longtable}
}
\longtab[3]{
\tiny
\def\arraystretch{0.75}
\begin{longtable}{lcccccc}
\caption{Members of ASCC 21a.}\\
\hline\hline
\textit{Gaia} DR2 id & $\alpha$ & $\delta$ & $\mu_\alpha\, \cos(\delta)$ & $\mu_\delta$ & $\varpi$ & $v_r$ \\
& $^\circ$ & $^\circ$ & $\mathrm{mas\, yr^{-1}}$ & $\mathrm{mas\, yr^{-1}}$ & $\mathrm{mas}$ & $\mathrm{km\, s^{-1}}$ \\
\hline
\endfirsthead
\caption{Continued.}\\
\hline\hline
\textit{Gaia} DR2 id & $\alpha$ & $\delta$ & $\mu_\alpha\, \cos(\delta)$ & $\mu_\delta$ & $\varpi$ & $v_r$ \\
& $^\circ$ & $^\circ$ & $\mathrm{mas\, yr^{-1}}$ & $\mathrm{mas\, yr^{-1}}$ & $\mathrm{mas}$ & $\mathrm{km\, s^{-1}}$ \\
\hline
\endhead
\hline
\endfoot
3220981762672284288 & 82.58990 & 0.11779 & 1.666$\pm$0.092 & -1.033$\pm$0.075 & 2.701$\pm$0.051 &  \\
3220974993803857024 & 82.54937 & 0.03798 & 1.603$\pm$0.192 & -0.907$\pm$0.167 & 2.690$\pm$0.105 &  \\
3221160845628179072 & 82.58878 & 0.75404 & 1.705$\pm$0.201 & -0.315$\pm$0.196 & 2.747$\pm$0.121 &  \\
3221174383365192832 & 83.11188 & 0.80872 & 1.558$\pm$0.067 & -0.748$\pm$0.064 & 2.853$\pm$0.035 &  \\
3221146242739617408 & 82.73343 & 0.69400 & 1.173$\pm$0.116 & -0.535$\pm$0.107 & 2.622$\pm$0.084 &  \\
3221133529636504960 & 82.70063 & 0.49666 & 1.422$\pm$0.207 & -1.136$\pm$0.161 & 2.730$\pm$0.142 &  \\
3221141054419085440 & 82.94693 & 0.61271 & 1.321$\pm$0.352 & -0.350$\pm$0.325 & 2.792$\pm$0.215 &  \\
3221146307163289728 & 82.79731 & 0.68081 & 1.414$\pm$0.152 & -0.756$\pm$0.133 & 2.840$\pm$0.139 &  \\
\hline
\end{longtable}
}
\longtab[4]{
\tiny
\def\arraystretch{0.75}
\begin{longtable}{lcccccc}
\caption{Members of ASCC 20.}\\
\hline\hline
\textit{Gaia} DR2 id & $\alpha$ & $\delta$ & $\mu_\alpha\, \cos(\delta)$ & $\mu_\delta$ & $\varpi$ & $v_r$ \\
& $^\circ$ & $^\circ$ & $\mathrm{mas\, yr^{-1}}$ & $\mathrm{mas\, yr^{-1}}$ & $\mathrm{mas}$ & $\mathrm{km\, s^{-1}}$ \\
\hline
\endfirsthead
\caption{Continued.}\\
\hline\hline
\textit{Gaia} DR2 id & $\alpha$ & $\delta$ & $\mu_\alpha\, \cos(\delta)$ & $\mu_\delta$ & $\varpi$ & $v_r$ \\
& $^\circ$ & $^\circ$ & $\mathrm{mas\, yr^{-1}}$ & $\mathrm{mas\, yr^{-1}}$ & $\mathrm{mas}$ & $\mathrm{km\, s^{-1}}$ \\
\hline
\endhead
\hline
\endfoot
3221011930522505472 & 81.59637 & 0.00374 & -0.297$\pm$0.059 & 0.102$\pm$0.048 & 2.769$\pm$0.037 &  \\
3221058518032896640 & 82.29722 & 0.56741 & -0.922$\pm$0.064 & 1.006$\pm$0.053 & 2.824$\pm$0.037 &  \\
3221057903852349824 & 82.22364 & 0.51868 & -0.161$\pm$0.180 & 0.932$\pm$0.143 & 2.848$\pm$0.085 &  \\
3221183797933580800 & 82.79538 & 0.84819 & -1.049$\pm$0.146 & 1.071$\pm$0.127 & 2.628$\pm$0.105 &  \\
3221191322716206848 & 82.95637 & 0.95098 & -0.850$\pm$0.041 & 1.092$\pm$0.037 & 2.704$\pm$0.026 &  \\
3221155275055026048 & 82.47784 & 0.71072 & -0.771$\pm$0.083 & 0.873$\pm$0.071 & 2.690$\pm$0.054 &  \\
3234169339495449344 & 80.07198 & 1.54039 & -0.283$\pm$0.105 & 0.944$\pm$0.089 & 2.822$\pm$0.070 &  \\
3234186588084188160 & 80.59184 & 1.65823 & -0.093$\pm$0.064 & 0.916$\pm$0.056 & 2.688$\pm$0.042 & 9.23$\pm$3.76 \\
\hline
\end{longtable}
}
\longtab[5]{
\tiny
\def\arraystretch{0.75}
\begin{longtable}{lcccccc}
\caption{Members of ASCC 18.}\\
\hline\hline
\textit{Gaia} DR2 id & $\alpha$ & $\delta$ & $\mu_\alpha\, \cos(\delta)$ & $\mu_\delta$ & $\varpi$ & $v_r$ \\
& $^\circ$ & $^\circ$ & $\mathrm{mas\, yr^{-1}}$ & $\mathrm{mas\, yr^{-1}}$ & $\mathrm{mas}$ & $\mathrm{km\, s^{-1}}$ \\
\hline
\endfirsthead
\caption{Continued.}\\
\hline\hline
\textit{Gaia} DR2 id & $\alpha$ & $\delta$ & $\mu_\alpha\, \cos(\delta)$ & $\mu_\delta$ & $\varpi$ & $v_r$ \\
& $^\circ$ & $^\circ$ & $\mathrm{mas\, yr^{-1}}$ & $\mathrm{mas\, yr^{-1}}$ & $\mathrm{mas}$ & $\mathrm{km\, s^{-1}}$ \\
\hline
\endhead
\hline
\endfoot
3220609337467853440 & 81.54291 & -0.37999 & 0.202$\pm$0.113 & 1.599$\pm$0.102 & 2.259$\pm$0.068 &  \\
3220550101279081600 & 81.67344 & -0.68123 & 0.039$\pm$0.065 & 0.056$\pm$0.058 & 2.368$\pm$0.040 &  \\
3220541099027590912 & 82.04015 & -0.68611 & -0.441$\pm$0.211 & 1.652$\pm$0.182 & 2.393$\pm$0.184 &  \\
3220580406568194432 & 81.30985 & -0.60245 & -0.322$\pm$0.125 & -0.001$\pm$0.111 & 2.417$\pm$0.081 &  \\
3220577108033309824 & 81.35058 & -0.63111 & -0.172$\pm$0.055 & 0.207$\pm$0.049 & 2.397$\pm$0.040 &  \\
3220555942434579072 & 81.76599 & -0.54059 & -0.108$\pm$0.147 & 1.495$\pm$0.135 & 2.383$\pm$0.078 &  \\
3220477808390090112 & 81.03160 & -0.81081 & -0.588$\pm$0.065 & 2.085$\pm$0.050 & 2.482$\pm$0.035 &  \\
3220636103704014208 & 81.51756 & -0.03916 & 0.168$\pm$0.312 & 0.976$\pm$0.260 & 2.540$\pm$0.199 &  \\
\hline
\end{longtable}
}
\end{appendix}





\end{document}